\documentstyle[12pt,epsfig]{article}

\newcommand{\be}{\begin{equation}}
\newcommand{\ee}{\end{equation}}
\begin{document}  
\topmargin 0pt
\oddsidemargin=-0.4truecm
\evensidemargin=-0.4truecm
\renewcommand{\thefootnote}{\fnsymbol{footnote}}
\newpage
\setcounter{page}{0}
\begin{titlepage}   
\vspace*{-2.0cm}  
\begin{flushright}
hep-ph/0510415
\end{flushright}
\vspace*{0.1cm}
\begin{center}
{\Large \bf $ ^7 Be$ Neutrino Signal Variation in KamLAND} \\ 
\vspace{1cm}

{\large 
Bhag C. Chauhan\footnote{On leave from Govt. Degree College, Karsog (H P) 
India 171304. E-mail: chauhan@cftp.ist.utl.pt}\\
\vspace{0.15cm}
{  {\small \sl CENTRO DE F\'{I}SICA TE\'{O}RICA DAS PART\'{I}CULAS (CFTP) \\
 Departamento de F\'\i sica, Instituto Superior T\'ecnico \\
Av. Rovisco Pais, P-1049-001 Lisboa, Portugal}\\
}}
\end{center}
\vglue 0.6truecm
\begin{abstract}
Large Mixing Angle (LMA) neutrino oscillation is the main solution for the long-standing Solar Neutrino Problem (SNP). Whether there is any subdominant effect accompanying the dominant LMA solution can not be ruled out at the moment, but will be settled by the forthcoming data from highly skilled real time experiments targeting essentially the low energy domain of solar neutrinos.
Assuming a subdominant effect converting one of the active neutrinos into a sterile partner in the varying solar field with changing sunspot activity, we performed field-profile-independent predictions for $^7 Be$ neutrino signal variation, which might be tested in the KamLAND's future solar neutrino detection program. We found that after a substantial reduction of background and running of KamLAND solar mode through the sunspot maximum period (around 2010 - 2012), when the solar field at the resonance may vary from few $kG$ to $300~kG$, the subdominant time variation effect might be clearly visible (more than $5\sigma$) for $^7 Be$ neutrinos.
\end{abstract}

\end{titlepage}   
\renewcommand{\thefootnote}{\arabic{footnote}}
\setcounter{footnote}{0}
\section{Introduction}
The KamLAND experiment \cite{KamL_1} seems to confirm LMA as a solution to the decades long puzzle of missing solar neutrinos. Although LMA has been accepted as the dominant solution to the problem, there are several open questions which still remain to be answered \cite{Cl_time}. 
Periodicity in the solar neutrino data, although controversial \cite{sk_sno, peter_david}, has been claimed recently by the Stanford Group \cite{peter_low, peter_high} upon examination of time binned data from all experiments except, so far, SNO. 
If it is confirmed, it will probably require Resonant Spin Flavor Precession (RSFP) within the sun affecting neutrinos in addition to the LMA effect. RSFP mechanism \cite{SFP} converts active neutrinos of one flavour into active or sterile (anti)neutrinos of another through the interaction of their magnetic moments with the transverse magnetic field on the way of their propagation. If the solar field is time varying this flavour conversion is also time dependent.  A significant time-modulation in the solar neutrino flux would imply the existence of a sizable neutrino magnetic moment $\mu_{\nu}$, and hence a wealth of new physics. 

Most of the solar neutrino data analyzed till date \cite{Ahmad:2002jz, lma_sno} is due to the high energy neutrinos (mostly E$~>~5~$MeV), and in this way the low energy solar neutrino domain (essentially $<~ 1-2 MeV$) is not much explored yet. 
In this sector the data we have so far are recorded by the inclusive radiochemical experiments Homestake, GALLEX/ GNO and SAGE. Examined in the time basis, some interesting signatures of time variability have been found in the data: The Stanford Group \cite{peter_low} claims a strong time variation effect for low energy neutrinos and, besides the unproven claim of time variation in Homestake Cl experiment \cite{Cl} \footnote{The long time variation suggested several years ago for Homestake Cl rate is still unclear \cite{Cl_time}.}, a time dependence in Ga data has been also noted \cite{Ga_rates}.
The present statistics of experiments are however not enough to establish this effect. So, in order to confirm or rule out a time dependence in low energy solar neutrino sector we need real time experiments and enough statistics.  

Owing to the hints of flux variability signature suggested by the Gallium and Homestake Cl experiments, and as both the experiments have a sizable contribution of $^7\!Be$ neutrino flux, it becomes worthwhile to study the $^7\!Be$ neutrino line separately. 
Time variation for $^7\!Be$ neutrinos and other low energy neutrinos in the real time experiments BOREXINO and LENS has been examined in \cite{jhep3}. Our main motivation in this work is, as indicated by the title itself, to give predictions for the $^7\!Be$ neutrino time variation in KamLAND experiment. KamLAND's future plan is to detect and analyze solar active neutrinos in the low energy sector with $^7\!Be$ neutrino measurements in the near future. 

The paper is organised as follows: in section 2 we discuss a possible time modulation of low energy solar neutrinos. The results and predictions for $^7\!Be$ neutrino time modulation in KamLAND are exhibited in section 3, and finally in section 4 the work is concluded.  

\section{Low Energy Solar Neutrinos ($LowE\odot\nu s$)}
Solar neutrinos are overwhelmingly represented by the low energy neutrinos: 99\% of total flux. The high energy tail (E$~>~5~$MeV) of solar neutrinos contributes less than $0.01$\% of the total flux.  
LMA has been established essentially for high energy neutrinos because most of the solar neutrino data registered in the several detectors till date are due to these \cite{Ahmad:2002jz, lma_sno}. 
On the other hand, $LowE\odot\nu s$, even though with dominating fraction, are almost concealed from the up to now studies of solar neutrinos. The quest for new physics in the $LowE\odot\nu$ sector is at present the major objective for solar neutrino studies.   

\subsection{Time Modulation in $LowE\odot\nu s$}
Apart from the unsettled claim of time variation in Homestake Cl experiment \cite{Cl, Cl_time} Gallium experiments show a discrepancy of $2.4\sigma$ between the combined results of two different periods of time \cite{Ga_rates}, which might be translated into a long time periodicity. Notice that Ga experiments have a dominant contribution of low energy neutrinos ($pp+^7\!Be~ \sim 80$\%). So, a strong time variation in Ga rates will definitely indicate time modulation in $LowE\odot\nu$ sector. 

In order to explain apparent signatures of time modulation in solar neutrino data \cite{peter_low, peter_high, Cl, Ga_rates} 
we developed a model \cite{jhep1}, which predicts a clear time variational effect in the low energy sector of solar neutrinos and a hardly visible effect in the high energy tail of solar neutrino spectrum. In other words, model forebodes a maximum time variation of Ga rates, moderate of Cl rates, and negligible effects on the SK and SNO rates. 
In the model, the magnitude of new mass squared difference ($\Delta m^2_{10}$) between one of the active states and the sterile one is chosen such that $LowE\odot\nu s$ have spin flavor precession resonating near the base of solar convective zone where the value of solar fields is peaked \cite{antia}. In this way there results a strong time modulation effect for these neutrinos. For qualitative illustrations see figure 1.  
\begin{figure}[h]
\setlength{\unitlength}{1cm}
\begin{center}
\hspace*{-1.6cm}
\epsfig{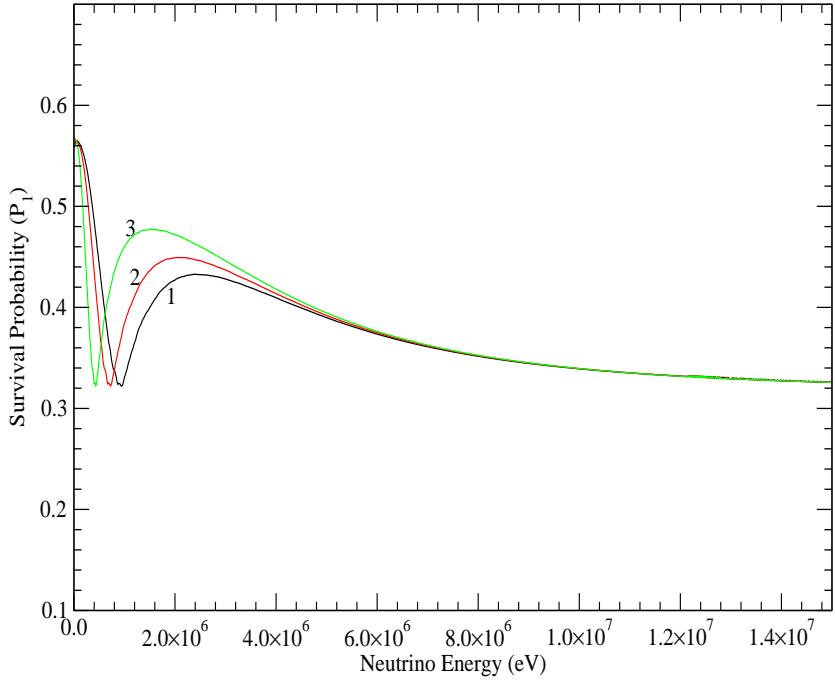}
\end{center}
\caption{ \it Time modulation effect in the low energy domain of solar neutrinos seen as dips (1,2,3) on survival probability curve due to the subdominant RSFP mechanism.} 
\label{fig1}
\end{figure}
To obtain this figure we used solar field profile 1 of \cite{jhep3} with an average field strength $\bar B_o=150~kG$. We plotted three probability curves showing strong suppression dips due to the subdominant RSFP mechanism in the low energy domain. The three dips (1,2,3) in the figure correspond to the three values of $\Delta m^2_{10}$, respectively as: \\
1. $-1.3\times10^{-8}~eV^2$; 2. $-1.0\times10^{-8}~eV^2 $; 3. $-0.6\times10^{-8}~eV^2$.

Time dependence of $LowE\odot\nu$ fluxes is now an open possibility which might be settled by the transparent measurements with the forthcoming real time neutrino detectors like BOREXINO \cite{Borexino}, LENS \cite{LENS}, and several others in the row \cite{real_time}. KamLAND, a reactor antineutrino detector, is also programing to detect solar neutrinos, and would substantially contribute towards settling of this issue. 

\section{$^7 Be$ Neutrino Time Modulations in KamLAND}
KamLAND is aiming at the physics of $LowE\odot\nu s$, in particular $^7\!Be$ neutrino line, by lowering its threshold and eliminating the troubling background due to other sources. 
It will measure $^7 Be$ signal via $\nu-e$ elastic scattering (ES): $$\nu~+~e~\rightarrow~\nu^{'}~+~e^{'}$$ 
where $(')$ signifies the same particle but with energy changed.

In order to perform predictions for $^7 Be$ neutrino time modulation in KamLAND, we estimated backgrounds and errors on the basis of informations collected from \cite{Junpei, U_Th}. For preventing the external gamma ray backgrounds from outside the liquid scintillator balloon we assumed its fiducial volume corresponding to a radius $R<4m$ leading to $7.2 \times10^{31}$ target electrons.
We assumed a substantial reduction of  $^{85} Kr~(\beta$-decays) background, although it is at present predominant in KamLAND over the $^7 Be$ neutrino signal region, and has to be reduced with a huge factor of order $10^6$. We also neglected the background due to $^{210} Po~(\alpha$- decays), although it needs to be suppressed by a factor of $\approx 10^5$, otherwise $^7 Be$ neutrino signal predictions become precarious essentially below 400 $keV$. We also assumed equilibrium after statistical subtraction of $^{238} U$, $^{232} Th$ backgrounds. By taking into account the smaller flux weights and negligible suppressions, we safely neglected the effect of time variations on the background due to $pep,~ ^{15} O$ and $^{13} N$ neutrinos. This assumption is perfect for the energy window 400 $keV$ to 800 $keV$. 

We used a Gaussian energy resolution function with $\sigma(E)= 6.2\%/\sqrt{E}$ as published by KamLAND collaboration in \cite{Araki}, and adopted BP05(OS) \cite{bahcall} SSM flux value for $^7 Be$ neutrinos.
Regarding the LMA parameters we considered KamLAND best fit for rate-and-shape analysis $\Delta m^2_{21}= 7.9^{+0.6}_{-0.5}\times 10^{-5}~eV^2$ and $tan^2\theta=0.46$ \cite{Araki}. We didn't use the combined data fit results of KamLAND + Solar Neutrino Data ($\Delta m^2_{21}= 7.9^{+0.6}_{-0.5}\times 10^{-5}~eV^2$ and $tan^2\theta=0.40^{+0.10}_{-0.07}$) because of the following reasons: 1) In the analysis of solar neutrino data only pure LMA has been considered and no subdominant RSFP effect is taken into account. 2) The values of the parameters we used are within $1\sigma$ of the combined KamLAND+Solar Analysis.  
Throughout the analysis we used $\mu_{\nu}=1\times 10^{-12}~\mu B$ and allowed the average peak field value $\bar B_o$ to vary in the range $10-300~kG$.
We chose the solar field profile 1 from \cite{jhep3}, as preferred by the solar data fits, peaked near the bottom of convective zone as suggested by the observations \cite{antia}. 

\subsection{$^7 Be$ Neutrino Spectrum and Model Predictions}
We studied the visible electron energy spectrum in the region from 280 $keV$ to 800 $keV$, which is essentially dominated by the smeared $^7 Be$ neutrino signal, once the optimum reduction of background is achieved.
The visible electron kinetic spectrum is calculated as
\be
R_{ES}=Q_o\int_{0}^{T^{'}_M}dT{'} f(T{'},T) \phi(E)
[P_1(E)\frac{d\sigma_{W}}{dT^{'}}+P_2(E))\frac{d\sigma_{W^{'}}}{dT{'}}].
\ee
Here $Q_o$ is the overall normalization factor, $\phi(E)$ is the SSM flux of $^7\!Be$ neutrinos and $E$ is their energy.
$P_1(E)$ is the survival probability of $\nu_e$, while $P_2(E)$ the production probability of $\nu_{\mu}$ such that $P_1(E)~+~P_2(E)\le 1$ \footnote{Equality sign holds for pure LMA, when there is no time modulation.}. 
The quantity $f(T^{'},T)$ is the energy resolution function of the detector in terms of the physical ($T^{'}$) and the measured ($T$) electron kinetic energy, and $\frac{d\sigma_W}{dT^{'}}$ and $\frac{d\sigma_{{W}{'}}}{dT^{'}}$ are the weak differential cross-sections for $\nu_{e}e$ and $\nu_{\mu} e$ scatterings, respectively.

Since $^7 Be$ neutrinos consist of a single energy line, by fixing $\Delta m^2_{10}$ all the neutrinos resonate exactly at the same position and experience the same field value. 
As a matter of fact, these neutrinos are blind to the shape of solar field profile, and in this way, because of this characteristic of $^7 Be$ neutrinos, the analysis is free from the uncertainties arising due to the infinite possibilities of field profile shape\footnote{The current knowledge about solar field profile is quite hazy; we have some observational support only for the peak field value $B_o$ \cite{antia}.}. 
In other words, the predictions for $^7 Be$ signal in KamLAND are quite robust in the sense that they are solar field-profile-independent.

In order to work out the maximum possible time variation of $^7 Be$ neutrinos we chose $\Delta m^2_{10}=-1.3\times10^{-8}$ so that $^7 Be$ neutrinos resonate near the peak of solar field. 
At the sunspot maximum the peak field value could be as high as $(300~-~500)~kG$ \cite{antia}. We performed predictions for the electron spectrum (events/keV) in KamLAND for 3 years of its continuous operation, as these will preserve the long time variational effects due to sunspot activity, if any.
In this way KamLAND could see a continuous increasing suppression up to the sunspot maximum (2010 - 2012) and a continuous decrease in suppression subsequently.\vspace{0.7cm}
\begin{figure}[h]
\setlength{\unitlength}{1cm}
\begin{center}
\hspace*{-1.6cm}
\epsfig{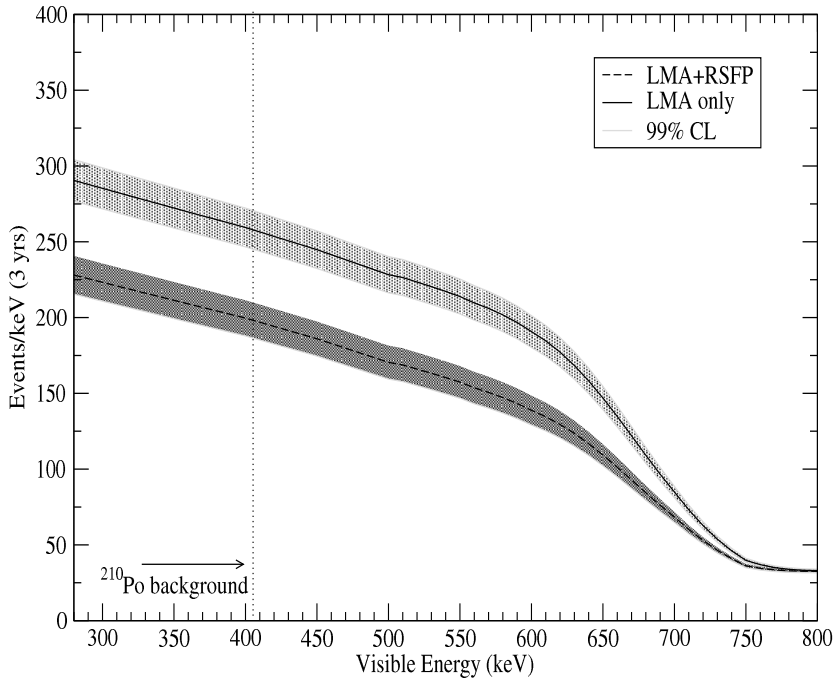}
\end{center}
\caption{ \it Predicted visible electron energy spectrum events/ keV in KamLAND continuously running for 3 years. Lower curve refers to the spectrum due to the maximum possible time modulation and the shaded area around the curves refers to 99\% CL.} 
\label{fig2}
\end{figure}

In figure 2 we have shown the expected spectrum in events per keV in the visible electron kinetic energy ranging from 280 $keV$ to 800 $keV$. The maximum possible suppression is shown by the lower curve with the average peak field $\bar B_o=300~kG$. 
The 99\% CL errors are shown by the shaded region around the curves. The predictions of these curves and errors will be modified on the left of the vertical dotted line, once the $^{210} Po$ background is taken into account. It is quite evident from the figure that in such a situation the time modulation effect during the solar cycle maximum (2010 - 2012) is clearly distinguishable (more than $5\sigma$) in KamLAND data-taking for 3 years. For weaker fields ($\bar B_o=140~kG$) the two effects (pure LMA and LMA+RSFP) might be still distinguishable at 99\% CL.

After neglecting $^{210} Po$ background and the time variation effects due to other solar neutrinos, which we did just for simplicity, the total number of events in the energy range 280 $keV$ to 800 $keV$ expected in KamLAND's three years of running is shown in table I. We have chosen four test values for the average peak fields $\bar B_o$. The first refers to pure LMA ($\bar B_o=0~kG$), and the others to LMA + RSFP. In particular, the third value refers to a field of $\bar B_o=140~kG$ (capable of separating time modulations from pure LMA oscillations at 99\% CL), and the last value $\bar B_o=300~kG$ refers to the maximum possible value of the peak field. 
  
\begin{center}
\begin{tabular}{ccccc} \\ \hline 
    & $280~keV$ & --- & $800~keV$ & \\
\hline
    & $\overline{B_o}$ & $^7 Be$ signal & Bkgd. & Total  \\
\hline
LMA  & $0~kG$ & $6.278\times10^4$ & $3.642\times10^4 $   & $9.920\times10^4$ \\
\hline
LMA+RSFP  & $100~kG$ & $5.926\times10^4$ & $3.642\times10^4$ & $9.568\times10^4$   \\
   & $140~kG$ & $5.618\times10^4$ & $3.642\times10^4$ & $9.261\times10^4$  \\
         & $300~kG$ & $3.961\times10^4$ & $3.642\times10^4$ & $7.603\times10^4$  \\
\hline
\hline
\end{tabular}
\end{center}

{\it{Table I - Total number of events in the energy range 280 $keV$ to 800 $keV$expected in 3 years of KamLAND data-taking.}}

In order to avoid the contaminations due to $^{210} Po$ and the significant time variation effects on the background due to $pep,~ ^{15} O$ and $^{13} N$ solar neutrinos, and to make our predictions more tangible, it is legitimate to confine our analysis inside the window 400 $keV$ - 800 $keV$. In this window KamLAND would only see the $^7 Be$ neutrino signal once the backgrounds due to $^{210} Bi$ and $^{85} Kr$ are significantly reduced. The total number of events expected in 3 years of KamLAND data-taking are shown below in table II. 
\begin{center}
\begin{tabular}{ccccc} \\ \hline 
   & $400~keV$ & --- & $800~keV$ & \\
\hline
     & $\overline{B_o}$ & $^7 Be$ signal & Bkgd. & Total  \\
\hline
LMA  & $0~kG$ & $4.285\times10^4$ & $3.015\times10^4$ & $7.301\times10^4$ \\
\hline
LMA+RSFP  & $100~kG$ & $4.046\times10^4$ & $3.015\times10^4$ & $7.061\times10^4$   \\
   & $140~kG$ & $3.836\times10^4$ & $3.015\times10^4$ & $6.851\times10^4$  \\
         & $300~kG$ & $2.705\times10^4$ & $3.015\times10^4$ & $5.720\times10^4$  \\
\hline
\hline
\end{tabular}
\end{center}

{\it{Table II - Total number of events in the energy range 400 $keV$ to 800 $keV$ expected in 3 years of KamLAND data-taking.}}

\vspace{0.7cm}
\begin{figure}[h]
\setlength{\unitlength}{1cm}
\begin{center}
\hspace*{-1.6cm}
\epsfig{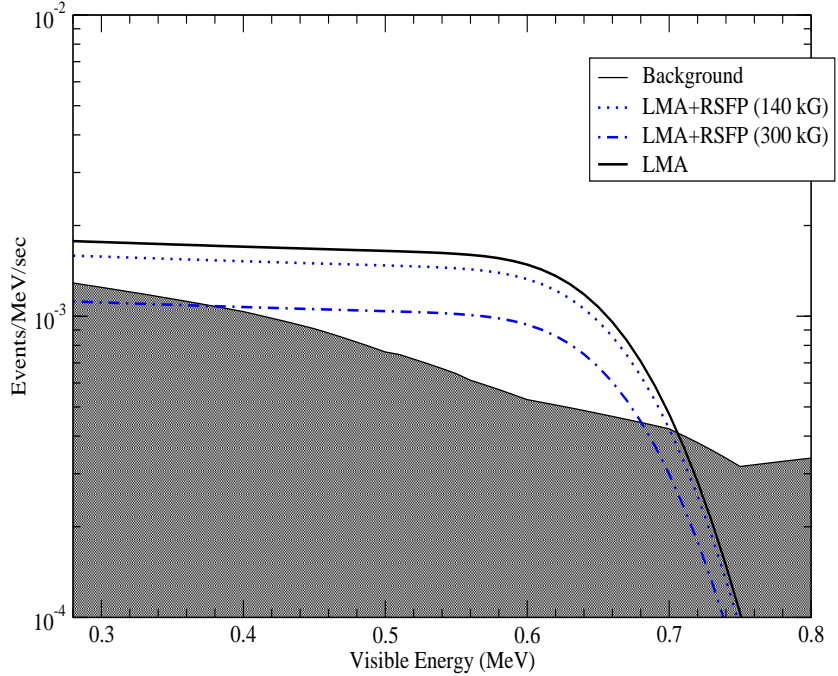}
\end{center}
\caption{ \it Individual $^7 Be$ and background visible electron energy spectrum events/ MeV/ sec in KamLAND. The dash-dotted spectrum refers to the maximum possible time modulation.} 
\label{fig3}
\end{figure}

The time modulation of individual $^7 Be$ neutrino spectrum in the energy range 280 $keV$ to 800 $keV$ is also shown in figure 3. The shaded area below the thin solid line as seen in the figure is the total background estimated in the KamLAND signal. The background is insensitive to the time modulation effect, which can be also concurred with the constant background event numbers given in the tables. However the $^7 Be$ neutrino signal is obviously suppressed as the solar cycle evolves. In the figure the lowermost blue dot-dashed curve corresponds to the $^7 Be$ neutrino signal when $B_o=300~kG$ (sunspot maximum), while the uppermost solid black corresponds to $B_o=(0~-~10)~kG$ (sunspot minimum, pure LMA). The blue dotted line refers to the suppression at 99\% CL. 

It is evident from the figure that the lowermost curve is almost disappearing in the huge shade of background, implying that the KamLAND visible electron spectrum shape is being dominated by that of background, and in this way it would be quite hard to make out the shape $^7 Be$ neutrino spectrum. In other words, during sunspot minimum, which KamLAND might see around 2006 - 2007, the visible spectrum shape will be close to that of $^7 Be$ neutrino spectrum, however at the sunspot maximum the KamLAND integrated spectrum shape might be tampered by the background. 
In the later case if the background is significantly reduced, the dominance of $^7 Be$ neutrino spectrum shape could be retrieved.
\vspace{0.7cm}
\begin{figure}[h]
\setlength{\unitlength}{1cm}
\begin{center}
\hspace*{-1.6cm}
\epsfig{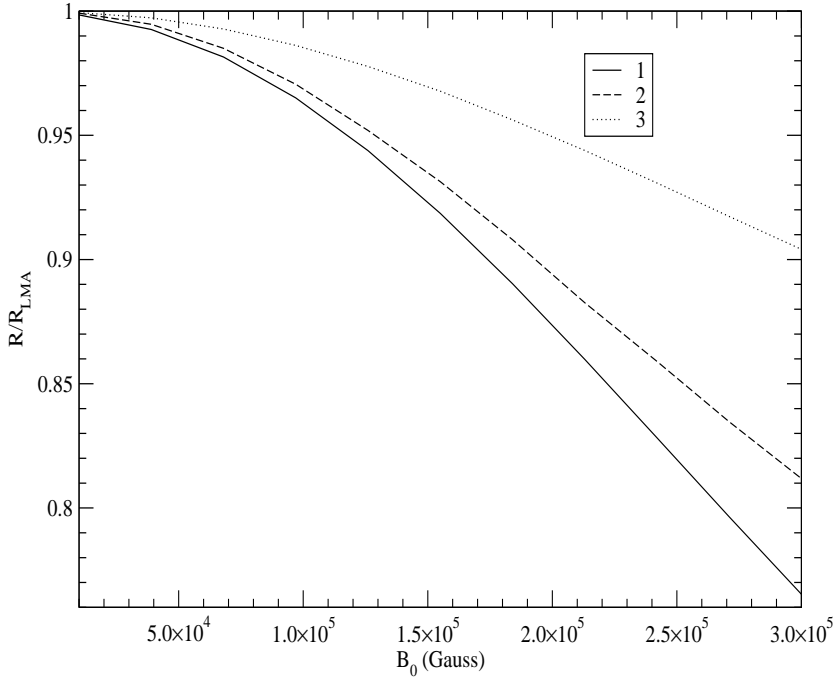}
\end{center}
\caption{ \it Time variation in KamLAND with changing solar fields shown for three values of $\Delta m^2_{10}$.} 
\label{fig4}
\end{figure}
In figure 4 we have depicted the total rate variation for three model values of $\Delta m^2_{10}$ with the changing solar peak field value, with reference to pure LMA rate. We recall that the pure LMA solution corresponds to the field $B_0=(0-10~kG)$ and, therefore, no modulations: $R/R_{LMA}=1$, as seen in the figure. Modulations might push the pure LMA rate down by $\sim 24\%$  as shown by curve 1 exhibiting the maximum sensitivity to solar activity. In case of curve 2 the rate decreases slower, but faster than the curve 3 for increasing $B_0$. It is evident that curve 3 shows the least sensitivity to the solar field because $^7 Be$ neutrinos resonate far outside the base of solar convective zone and do not see an adequate field for their non-adiabatic conversion into sterile neutrinos. However, in this case $pp$ neutrinos, having their resonance deeper inside sun around the peak field zone, may suffer maximum conversion into steriles, an effect which is beyond the current scope of KamLAND.  

\section{Conclusions}
After an independent test confirmation of neutrino oscillation performed by KamLAND over a short baseline of $\sim 200$ kms, LMA has been accepted as a dominant solution to the SNP. Up to now most of the solar neutrino data are provided by the high precision real time experiments (SNO and SK), which have been monitoring the high energy tail of the solar neutrino spectrum. However, the dominant solar neutrino flux (99\%), which is due to low energy sector, is almost unexplored. 
On the other hand, the Stanford Group \cite{peter_low} claims even stronger time variation effects for $LowE\odot\nu s$.
So, any subdominant effect in the low energy sector accompanying LMA can not be ruled out at the moment, because we don't have enough data and real time measurements which could establish or rule out any such effect. 
Nevertheless, such time variations, if they exist, could be certainly tested by the forthcoming high statistics real time $LowE\odot\nu$ experiments \cite{real_time}.
 
In this work we performed predictions for $^7 Be$ neutrino signal variation in KamLAND, assuming a possible correlation with sunspot activity. We assumed the $^7 Be$ neutrinos to resonate near the peak of solar field in order to examine the maximum possible time-variation in KamLAND. The results are quite robust for the reason that they are solar field-profile-independent.  
We have found that the time modulation effect might be clearly distinguishable from the pure LMA solution  (more than $5\sigma$) in KamLAND data-taking for 3 years essentially corresponding to the sunspot maximum period around 2010 - 2012. It has also been seen that an average field of $\bar B_o=140~kG$ can produce an effect capable of separating time modulations from pure LMA at 99\% CL. At the time of low sunspot activity around 2006 - 2008, the KamLAND electron spectrum shape is expected to be close to that of $^7 Be$ neutrino spectrum, and at the sunspot maximum the KamLAND integrated spectrum shape will be tampered by the background. Once the background is significantly reduced however, dominance of $^7 Be$ neutrino spectrum shape can be retrieved.
Finally, we have also noted that at the time of sunspot maximum, the $^7 Be$ neutrino event rate might be further suppressed by $\sim 24\%$ relative to pure LMA.   

\vspace{1cm}
\noindent {\Large \it Acknowledgments}\\
{\em Author is grateful to Junpei Shirai from the KamLAND Collaboration for providing relevant informations about the detector, and Jo\~{a}o Pulido for useful suggestions and a careful reading of the manuscript. The work was supported by Funda\c{c}\~{a}o para a Ci\^{e}ncia e a Tecnologia through the grant SFRH/BPD/5719/2001.}


\end{document}